\documentclass[12pt, aps,preprintnumbers,amsmath,amssymb,nofootinbib]{revtex4}
\pdfoutput=1
\usepackage{graphicx}
\usepackage{cancel}
\usepackage{amssymb}
\usepackage{amsmath,dsfont,textcomp}
\usepackage{bm}
\usepackage{color}
\usepackage{euscript,amsmath}

\def\dslash{\rlap{\hspace{0.02cm}/}{\partial}}

\usepackage{graphicx}% Include figure files
\usepackage{dcolumn}% Align table columns on decimal point
\usepackage{bm}% bold math
\usepackage{epsf}
\usepackage{amssymb,amsmath,amsfonts,amsbsy}%,epsfig}%,psfrag,axodraw}
\usepackage{color}
\usepackage{fancyhdr}
\def\lsim{\mathrel{\rlap{\lower4pt\hbox{\hskip1pt$\sim$}}
    \raise1pt\hbox{$<$}}}         %less than or approx. symbol
\def\gsim{\mathrel{\rlap{\lower4pt\hbox{\hskip1pt$\sim$}}
    \raise1pt\hbox{$>$}}}
\def\fun#1#2{\lower3.6pt\vbox{\baselineskip0pt\lineskip.9pt
  \ialign{$\mathsurround=0pt#1\hfil##\hfil$\crcr#2\crcr\sim\crcr}}}

\input epsf

\newcommand{\eV}{~{\rm eV}}
\newcommand{\GeV}{~{\rm GeV}}
\newcommand{\TeV}{~{\rm TeV}}

\begin{document}

\draft
\preprint{\tt IPMU10-0149, \,\,\, \\ \today}

\title{Ghost condensation and {\it CPT} violation in neutrino sector}

\vspace{1.0cm}

\author{Shinji Mukohyama}
\email{shinji.mukohyama@ipmu.jp}
\author{Seong Chan Park}
\email{seongchan.park@ipmu.jp}
\vspace{1cm}
\affiliation{Institute for the Physics and Mathematics of the Universe, The University of Tokyo, Chiba $277-8568$, Japan
}

\vspace{2.0cm}

\begin{abstract}
Motivated by the recent MINOS observation, $\Delta m_{32}^2\neq \Delta \bar{m}_{32}^2$, we study {\it CPT} violation  in neutrino sector caused by ghost condensation.  As a concrete model, we consider an extra dimension model with two branes, which are spatially separated and communicate only through singlet `messenger' fermion fields which contain right handed neutrinos in the bulk. Assuming ghost condensation on one brane and localization of the standard model sector on the other brane, {\it CPT} violation occurs only in neutrino sector by Yukawa couplings with the messengers, which, at the same time, lead small neutrino masses. In the parameter space which is consistent with the MINOS observation  suggests possible observation of the twinkling cosmic microwave background radiation and blinking light-ray from a quasar at a large distance, with a timescale about $0.1$ day. 
\end{abstract}

 %(see also 11.25.Mj Compactification and four-dimensional models, 11.25.Uv D % branes)

\maketitle
%\tableofcontents

\newpage

\section{Introduction}

Recently preliminary result from MINOS was released \cite{minos}  \footnote{No sidereal modulation is found in the MINOS far detector \cite{minos 2}, which is expected as a signal of Lorentz and {\it CPT} violation in an extension of the SM (SME) framework \cite{sme}.
 We follow the conventional notations for mass differences and mixing angles, $\Delta m_{32}^2, \theta_{23}$ for neutrinos and  $\Delta \overline{m}_{32}^2, \bar{\theta}_{23}$ for antineutrinos, respectively.}: 

\begin{eqnarray}
 &&\Delta m_{32}^2 = 2.35 \begin{matrix}
+0.11\\ 
-0.08
\end{matrix} \times 10^{-3} \eV^{2}, \,\nonumber \\
&&\Delta \overline{m}_{32}^2= 3.36  \begin{matrix}
+0.45\\ 
-0.40
\end{matrix}\times 10^{-3} \eV^{2},\,\\
&&\sin^2(2\theta_{23}) > 0.91 \nonumber \\
 &&\sin^2 (2 \bar{\theta}_{23})=0.86\pm 0.11 \nonumber
\label{eq:minos}
\end{eqnarray}
 where apparent difference between particle and anti-particle has been observed with 90$\%$ C.L  \cite{lykken}. This result, even though it is preliminary, is surprising since {\it CPT} theorem does not allow the difference between particle and anti-partilce in the well established frame work of local relativistic quantum field theory \footnote{See e.g., Ref. \cite{nelson} for a trial to understand the data without {\it CPT} violation}.  Indeed every known particle except neutrinos has the same mass as the {\it CPT} conjugate anti-particle within experimental resolutions. We take this observation as a chance to consider $CPT$ violation in neutrino sector.

%Figure from MINOS homepage
%\begin{figure}[b]
%      \includegraphics[width=.47\textwidth]{minos.pdf}
%  \caption{MINOS data released at Neutrino 2010 \cite{minos}. }
 % \label{fig:minos}
%\end{figure}
%

{\it CPT} violation implies violation of Lorentz invariance \cite{cpt}.
As a possible source of Lorentz violation, in the present paper we consider
ghost condensation~\cite{gc1}. Ghost condensation is a 
mechanism for spontaneous Lorentz breaking and leads to the simplest
Higgs phase of gravity. It is indeed the simplest in the sense that the 
number of associated Numbu-Goldstone boson is just one. The order
parameter of the spontaneous Lorentz breaking is the vev of derivative
of a scalar field $\langle\partial_{\mu}\phi\rangle$, which is supposed
to be timelike. By properly choosing the time coordinate, we can set
$\langle\partial_i\phi\rangle=0$ and, thus, ghost condensation is
characterized by a scale
$M\equiv\sqrt{|\langle\partial_0\phi\rangle|}$. As in the usual Higgs
mechanism, ghost condensation modifies infrared behavior of the force
law, in this case gravity. In the limit $M\to 0$, Lorentz symmetry and
thus general relativity are safely recovered. For this reason, there is
no phenomenological lower bound on $M$, unless additional assumptions
are made.  (See \cite{gc3} for an example of such an
additional assumption and the corresponding lower bound on $M$.) On
the other hand, there are universal upper bounds on $M$ and the strongest is
$M< 100\GeV$~\cite{gc2}.

There is an immediate question \cite{hitoshi}: why {\it CPT} violation is seen only in neutrino sector. There are various stringent constraints on {\it CPT}
violation in other sectors. The most severe one is from $K^0-\bar{K}^0$ mixing where $ (m_{K^0} -m_{\bar{K}^0})/m_{average} < 8\times 10^{-18}$
at CL=$90\%$ \cite{pdg, Murayama:2003zw}. In lepton sector, the bound is less severe but still significant as
$(m_{e^+}-m_{e^-})/m_{average}<8 \times 10^{-9}$ at CL=$90\%$
\cite{pdg}. Taking all these severe constraints into account, it is suggested that only neutrinos may directly couple to ghost condensation.  We notice that this situation is naturally realized in 
well motivated models where singlet neutrinos propagate in the bulk of extra dimension and the standard model particles are confined on a brane. The setup was originally suggested to explain small neutrino mass (see e.g., Refs.\cite{nima, neubert, park}). Now, in addition to the original setup, we suggest that ghost condensation takes place in a  brane which is spatially separate from the standard model brane.  In this case, the singlet neutrinos in the bulk naturally play the role of messenger fields through which the standard model sector communicates with ghost condensation. Then it is natural that a sizable {\it CPT} violation can occur only in neutrino sector. We will explicitly suggest a minimal model where all the required properties  are realized. 

The paper is organized as follows. In Sec.\ref{sec:model} we describe the model in 5D and derive the 4D effective action where the neutrino and anti-neutrino masses are calculated. In Sec. \ref{sec:cosmology} we consider various constraints on the model from cosmology and possible detection of ``twinkling CMB" in the future. We then summarize the paper in Sec. \ref{sec:conclusion}.

\section{Model}
\label{sec:model}

The model extra dimension is a flat\footnote{The setup can be built on warped extra dimension\cite{rs} as in \cite{neubert}.}  orbifold $S^1/Z_2$  with radius $R$  which can be equivalently described by an interval $I=[0, \pi R]$ where $y=0$ and $y=\pi R$ correspond to the fixed points of the orbifold.  We assume that the standard model sector is localized at $y=\pi R$ and  ghost condensation takes place at $y=0$. Singlet fermions, which contain right-handed neutrinos, are in the bulk so that they can mediate ghost condensation and the standard model sector. As a singlet does not couple to the standard model gauge bosons, it can naturally propagate through the bulk as graviton does. Essentially the present setup extends the one in \cite{nima} by introducing ghost condensation so that we can enjoy all the advantages of previous studies and allow the necessary $CPT$ violation.

\begin{figure}[t]
    \includegraphics[width=.7\textwidth]{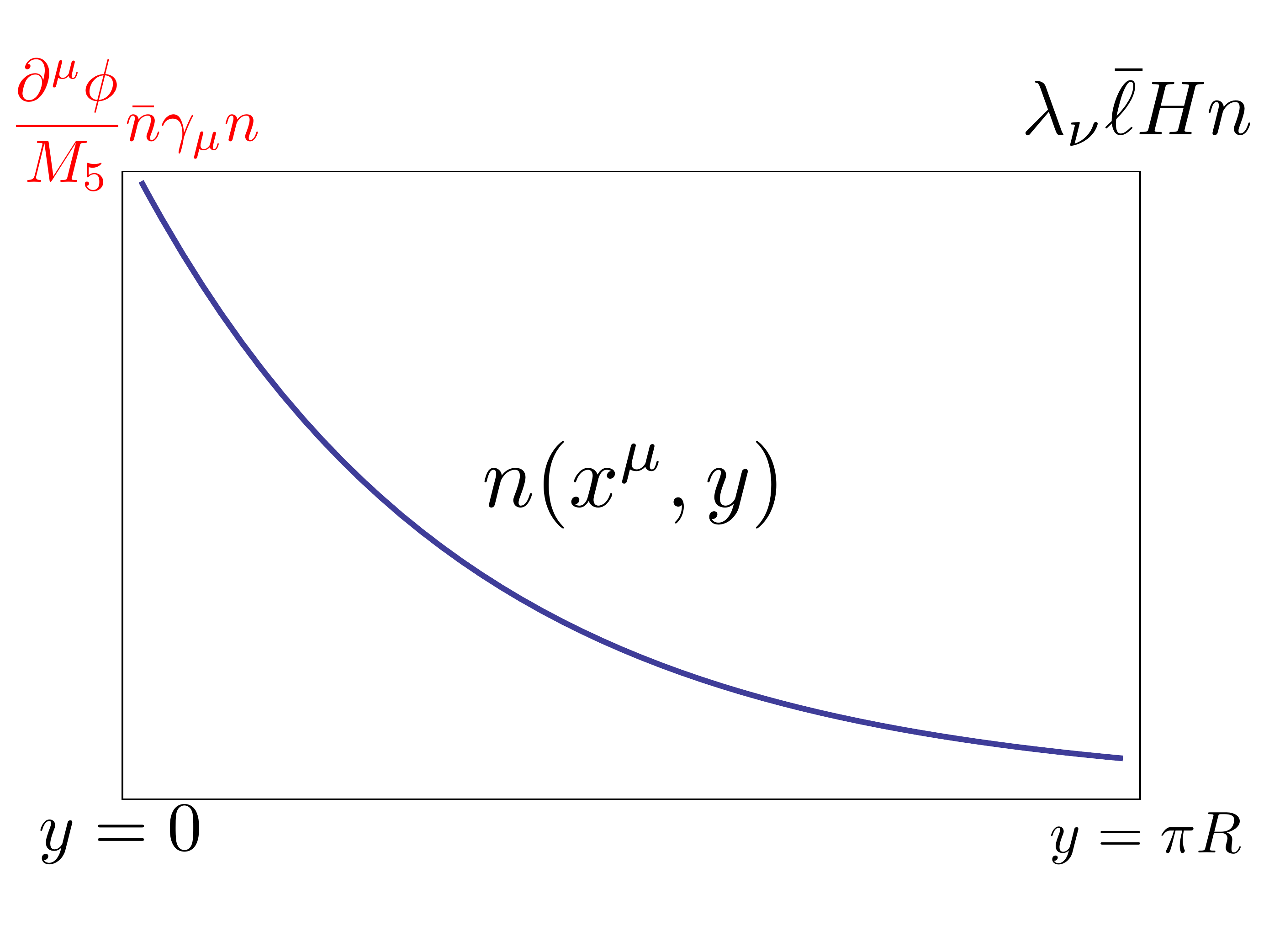}
  \caption{Our model.} 
  \label{fig:model}
\end{figure}

 The action is given as:
\begin{eqnarray}
S = \int d^4x \int_0^{\pi R} dy \left[ \tfrac{1}{\pi R}{\cal L}_{bulk} + \delta(y){\cal L}_0 + \delta(y-\pi R){\cal L}_\pi\right]
\end{eqnarray}
where ${\cal L}_\pi$ contains all the SM particles, i.e., quarks, leptons, gauge bosons of $SU(3)_c\times SU(2)_W \times U(1)_Y$ and interactions among them, ${\cal L}_0$ has ghost condensation so that Lorentz violation takes place. The bulk Lagrangian ${\cal L}_{bulk}$ contains singlet neutrinos, $n(x^\mu, y)$,  propagating in the bulk, $0\leq y \leq \pi R$ (See Fig. \ref{fig:model}). More explicitly,
\begin{eqnarray}
{\cal L}_{bulk}&=& i \bar{n}\Gamma^M \partial_M n - m~ \bar{n}n  +h.c.,\\
{\cal L}_0 &\ni& \frac{\partial_\mu \phi}{M_5}\bar{n} \gamma^\mu n + g.c., \\
{\cal L}_\pi &\ni& \lambda_\nu \bar{\ell}H n +h.c.+\cdots,
\end{eqnarray}
where $g.c.$ describes the dynamics of ghost condensation, $\Gamma^M =(\gamma^\mu, i \gamma^5)$ and `$\cdots$' constains the rest of SM sector and $M_5$ is Gravity scale in five dimension which is related to Planck scale as $M_{Pl}^2 = M_5^3 \pi R$. The bulk mass $m$ is consistent with orbifold symmetry as it is $m \propto {\rm sign}(y)$ \footnote{A double kink mass can be introduced so that Kaluza-Klein parity is conserved \cite{sued} but we do not pursue this direction here.}. 

To get the four dimensional effective action, we do the Kaluza-Klein(KK) decomposition as
\begin{eqnarray}
n(x,y) = \sum_k n_L^k(x) f_L^k(y) + n_R^k f_R^k(y),
\end{eqnarray}
where $f_L^k$ and $f_R^k$ are the dimensionless KK basis functions  satisfying the equation of motion:
\begin{eqnarray}
m_k f_{L/R}^k \mp \partial_y f_{R/L}^k - m f_{R/L}^k &=&0. 
%m_k f_R^k + \partial_y f_L^k -m f_L^k &=&0.
\label{eq:emo}
\end{eqnarray}

For zero mode, $k=0$, with $m_k=0$, the Eq. \eqref{eq:emo} is separable
\begin{eqnarray}
\partial_y f_{R/L}^0 \pm m f_{R/L}^0 &=&0, 
%\partial_y f_L^0 -m f_L^0 &=&0,
\end{eqnarray}
so is easily solved as
\begin{eqnarray}
f_{R/L}^0(y) &=& C_{R/L} e^{\mp m y}.
%f_L^0(y) &=& C_L e^{m y},
\end{eqnarray}
For $m>0$, right-handed zero mode is localized toward $y=0$ and left-handed zero mode is localized toward the other end point, $y=\pi R$. Here the normalization constant $C_{R/L}$ for $R/L$-handed zero mode are determined by the normalization condition $\int_0^{\pi R} |f_{R/L}^0|^2 dy =1$:
\begin{eqnarray}
C_{R/L} =\left( \frac{\pm 2 m \pi R}{1-e^{\mp2 m \pi R}}\right)^{1/2}.
%C_L = \left(\frac{2mL}{-1+e^{2 m L}}\right)^{1/2}.
\end{eqnarray}

If Dirichlet boundary conditions to $n_L$ is imposed, $f_L^k(0)=0=f_L^k(\pi R)$ at boundaries and right-handed states $n_R^k$ survive at boundaries and can interact with boundary localized fields \footnote{The five dimensional Lorentz invariance determines the relative sign of $Z_2$-parity of $n_L$ and $n_R$ or equivalently the boundary conditions to them. If odd-parity is imposed to one chiral state (or satisfying Dirichelt boundary condition), the other state has to have opposite parity, i.e., even parity (or satisfying (modified) Neumann boundary condition). }.

 %Fig
\begin{eqnarray}
S_{eff}&=&\sum_{k=0}^\infty i \overline{n_R^k}\dslash  n_R^k - m_k \overline{n_R^k} n_L^k +(R\leftrightarrow L)  \nonumber \\
 &+&
 \sum_k \lambda_\nu f_R^k(\pi R) \bar{\ell}H n_R^k  + h.c.\nonumber \\
&+&\sum_{k, \ell} \frac{\partial_\mu \phi}{M_5}\overline{n_R^k}\gamma^\mu n_R^\ell f_R^{k *}(0)f_R^\ell(0) +h.c. \\
&+&g.c. \nonumber 
%&\ni& {\cal M} \,\overline{{\cal N}_L}{\cal N}_R +h.c \nonumber \\
%&+& C_R^2 \frac{\partial_\mu\phi}{M_5} \overline{\nu_R}\gamma^\mu\nu_R
\end{eqnarray}
where $m_0=0$, $m_{k\geq1}=\sqrt{m^2+ (k/R)^2}$.  Using the collective notations for the left-handed and right-handed states, ${\cal N}_L = (\nu_L,n_L^1, n_L^2, n_L^3, \cdots)$ and ${\cal N}_R = (n_R^0, n_R^1, n_R^2, n_R^3, \cdots)$, respectively, we can construct the mass term $\overline{{\cal N}_L}\,{\cal M} \,{\cal N}_R +h.c. $ with $(N \times N)|_{N\to \infty}$ mass matrix 
\begin{eqnarray}
{\cal M} = \begin{pmatrix}
 \lambda_0 v  &  \lambda_1 v & \lambda_2 v & \lambda_3 v&\cdots & \lambda_N v\\ 
 0& m_1 & 0  & 0 & \cdots & 0\\ 
 0& 0 & m_2 & 0  & \cdots & 0\\ 
 0&  0& 0 & m_3 & \cdots & 0 \\ 
 \vdots & \vdots & \vdots & \vdots & \ddots & 0 \\
 0 & 0 & 0& 0 &0 & m_N
\end{pmatrix},
\end{eqnarray}
where $v = \langle H \rangle$ after electroweak symmetry breaking and
the effective Yukawa coupling for the n-th KK state is $\lambda_n =
\lambda_\nu f_R^n(\pi R)$. The physical mass eigenstates and their
masses could be obtained after diagonalizing ${\cal M}{\cal M}^\dagger
\to U^\dagger{\cal M}{\cal M}^\dagger U$ by ${\cal N}_L \to
U^\dagger{\cal N}_L$. If $\lambda_0 \equiv \lambda_\nu f_R^0(\pi R)$
were zero, the matrix  ${\cal M}{\cal M}^\dagger$ would have a zero
eigenvalue with the corresponding eigenvector 
\begin{equation}
 u_0 = \cos\theta_{\nu} \, \begin{pmatrix}
1 \\
-\lambda_1 v/m_1 \\
-\lambda_2 v/m_2 \\
\vdots \\
-\lambda_N v/m_N
\end{pmatrix}, \quad
\tan^2 \theta_\nu = \sum_{n=1} \frac{ \lambda_n^2 v^2}{m_n^2},
  \label{eq:mix}
\end{equation}
where the angle $\theta_\nu$ describes the mixing between the zero mode
and the higher KK modes. Thus, to the leading order in the small parameter 
$\lambda_0\ll 1$, we can get the lowest eigenvalue which gives the
physical neutrino mass squared: 
\begin{eqnarray}
 m_\nu^2 = u_0^{\dagger}\, 
  {\cal M}{\cal M}^\dagger\,  u_0
  = \lambda_0^2  v^2 \cos^2 \theta_\nu. 
\end{eqnarray}
 As the wave function of R-handed state at
 $y=\pi R$ is exponentially suppressed as $f_R^0(\pi R) \approx
 \sqrt{2m\pi R}e^{-m\pi R}\ll 1$, we naturally get a small neutrino
 mass, $m_\nu \sim 0.1 \eV$, with a largish $m R \simeq 10$ and
 $\lambda_\nu \sim 1$.  As it is clearly seen in Eq.\eqref{eq:mix},  the
 zero mode can mix with KK excitation modes but the mixing angle is
 highly suppressed $~{\cal O}(\lambda_\nu v /m_{KK})$. However the mass
 splitting between particle and antiparticle in the zero mode states,
 which we can take as a perturbation,  cannot be neglected and actually
 is important to account the MINOS anomaly. 
  
Once Ghost condensation takes place on $y=0$ brane, $\langle \partial_0 \phi \rangle\neq 0$,
the second term becomes another neutrino bilinear term
\begin{eqnarray}
C_R^2 \frac{M^2}{M_5} \overline{\nu_R} \gamma^0 \nu_R,
 \end{eqnarray}
where $M^2 = \langle \partial_0 \phi \rangle$. A convenient parameter, $\epsilon \equiv C_R^2 M^2/{M_5}$, measures the Lorentz violation. When $\epsilon \neq 0$  the masses of particle (neutrino, in our case) and anti-particle (anti-neutrino, in our case) are different by $\epsilon$ \footnote{The easiest way to see this difference is as follows. With the nonzero $\epsilon\neq 0$, the Dirac equation in momentum space is $(\gamma^\mu p_\mu -m_0 -\epsilon \gamma^0)\psi(p)=0$ where $m_0$ is the Dirac mass coming from conventional Higgs mechanism. In the rest frame, $p^\mu=(m, \vec{0})$, the equation is reduced to $ \left((m-\epsilon)\gamma^0 -m_0 \right)\psi(p)=0$ and the mass eigenvalues are $m = \pm m_0 +\epsilon$ or $|m|=|m_0 \pm \epsilon|$.}: 
\begin{eqnarray}
 m_{\nu}-m_{\bar{\nu}} \simeq \epsilon.
\end{eqnarray}
The MINOS data in Eq.\eqref{eq:minos} implies that the mass difference in neutrino and anti-neutrino would be in the range of $0.01 \eV$  or
\begin{eqnarray}
&\epsilon &= C_R^2 \times \frac{M^2}{M_5} \sim 0.01 \eV, \nonumber \\ 
&\Rightarrow& M\sim 15\GeV \times \left(\frac{R^{-1}}{1\TeV}\right)^{\tfrac{1}{6}}\times \left(\frac{\epsilon}{0.01\eV}\right)^{\tfrac{1}{2}}.
\end{eqnarray}
Here we used $C_R^2 \simeq 2 m \pi R \sim 60$ to fit the correct neutrino mass scale with $\lambda_\nu\sim 1$. Note that $\epsilon\sim 0.01\eV$ is
consistent with the bound $\epsilon<10^3\eV$ from neutron $\beta$
decay~\cite{Di Grezia:2005qf}. Again as the splitting is small enough we can neglect the mixing between zero mode and other KK modes.

\section{Implications to cosmology}
\label{sec:cosmology}

Taking $\epsilon \sim 0.01 \eV$ and the bound on the size of extra dimension\footnote{Since the right-haned singlet neutrino does not couple to the gauge bosons, it can  only be probed by processes involving Higgs field. As Yukawa coupling for KK-neutrino is not small, KK neutrinos can contribute to the Higgs decay ($H \to \ell n^k$) if they are light enough. Taking current mass bound for the Higgs into account, we conclude that $m_1  \gsim 100 \GeV$ or $R^{-1} \gsim 10 \GeV$  is allowed by LEP-II and Tevatron experiments.}, we get the lower
bound on $M$ from the neutrino sector, $M\gsim 10$ GeV. On the other hand, from
nonlinear dynamics of the gravity sector, we obtain the upper bound on
$M$~\cite{gc2}. Thus, there is only a small window for $M$:%
\begin{eqnarray}
10 \GeV \lsim M \lsim 100\GeV.
\label{eqn:boundsonM}
\end{eqnarray}

Since $M$ should be pretty close to $100\GeV$ we expect to see twinkling of the
CMB~\cite{gc2} and blinking of the light from a quasar at a large distance with the twinkling timescale  
\begin{eqnarray}
 T_{\rm twinkling} &\sim& \frac{M}{100\GeV}\cdot
  \frac{300km/s}{v}\times 0.1{\rm day},
\end{eqnarray}
where $v$ is the overall velocity of the rest frame of ghost
condensation relative to the CMB and the quasar rest frame, respectively.

There are other bounds which we should take into account but, as shall
see, they do not set additional constraints. Due to the coupling of
neutrinos to ghost condensation, an ultrarelativistic neutrino emits a
Nambu-Goldstone boson~\cite{ArkaniHamed:2004ar}. This is an analogue of
\v{C}erenkov radiation. Demanding that neutrinos from SN1987A should not
be deflected by this process, we obtain the bound~\cite{Grossman:2005ej} 
\begin{eqnarray}
 \epsilon < 10^4\eV \times \left(\frac{M}{100\GeV}\right)^{\tfrac{3}{2}}
  \times \left(\frac{0.1\eV}{m}\right).
\end{eqnarray}
This is automatically satisfied if (\ref{eqn:boundsonM}) is
satisfied. If we demand that the total relative energy lost from
cosmological neutrinos to Nambu-Goldstone bosons during the epoch
between neutrino decoupling to the last scattering surface of the CMB,
then we obtain the bound~\cite{Grossman:2005ej}  
\begin{eqnarray}
 \epsilon < 10^{9}\eV \times \left(\frac{M}{100\GeV}\right)^2
  \times \left(\frac{0.1\eV}{m}\right).
  \label{eqn:CMBbound}
\end{eqnarray}
Again, this is satisfied under (\ref{eqn:boundsonM}).

\section{Conclusion}
\label{sec:conclusion}
In this paper we have considered ghost condensation as a source of {\it CPT} violation. Taking the coupling with neutrino current into account, we could provide a possible difference between neutrino and anti-neutrino, namely $\Delta m^2 \neq \Delta \bar{m}^2$ which has been recently hinted by MINOS experiment even though the result is preliminary. To avoid a sizable {\it CPT} violation in other sectors, we suggest an extra dimension model where the right-handed neutrinos in the bulk mediate the standard model sector and the ghost condensate which takes place at a distance from the standard model brane.  The model leads to interesting observational consequences: twinkling CMB and blinking quasars with the timescale around tens to hundreds of minutes which might be within the reach of future CMB observations (e.g. Planck \cite{bluebook}) and astronomical point source observations.

Finally discussion on flavor non-diagonal {\it CPT} violation is in order. Flavor non-diagnoal dimension five operators directly contribute to neutrino oscillation as in the standard matter effects and so are highly constrained by solar, atmospheric and other ground based neutrino experiments.  To forbid flavor non-diagonal terms we assume flavor symmetry on the ghost condensate brane so that only flavor diagonal dimension five operators are induced at the leading order. On the other hand,  flavor transition induced by Yukawa couplings are small $\epsilon_{i\neq j} \sim \sum_k y_{ik} y_{kj}^* |f_R^0(\pi R)|^2 \epsilon \sim 10^{-24}\epsilon$.

%\vspace{0.5cm}
\begin{center}
{\bf Acknowledgement}
%\vspace{0.0cm}
\end{center}
The authors would thank Tsutomu Yanagida, Hitoshi Murayama and Masahiro Tanaka for helpful discussions and comments. 
This work was supported by the World Premier International Research Center Initiative 
(WPI initiative) by MEXT. The work of S.M. was supported in part by Grant-in-Aid for Young Scientists (B) No. 17740134, Grant-in-Aid for Creative Scientific Research No. 19GS0219, Grant-in-Aid for Scientific Research on Innovative Areas No. 21111006, Grant-in-Aid for Scientific Research (C) No. 21540278, and the Mitsubishi Foundation. S.C.P. is supported by the Grant-in-Aid for scientific 
research (Young Scientists (B) No. 21740172) from JSPS, Japan.


\begin{thebibliography}{999}

\bibitem{minos}
P. Vahle, for the MINOS collaboration,``New results from MINOS,"
talk at Neutrino 2010,  June 14-19, Athens, Greece (http://www.neutrino2010.gr/index.php).
Also see the official MINOS webpage (http://www-numi.fnal.gov/) where all the results could be found.

\bibitem{minos 2}
  P.~Adamson {\it et al.}  [MINOS Collaboration],
  ``A Search for Lorentz Invariance and CPT Violation with the MINOS Far
  Detector,''
  arXiv:1007.2791 [hep-ex].  
  

%\cite{Colladay:1996iz}
\bibitem{sme}
  D.~Colladay and V.~A.~Kostelecky,
  ``CPT violation and the standard model,''
  Phys.\ Rev.\  D {\bf 55}, 6760 (1997)
  [arXiv:hep-ph/9703464].
  %%CITATION = PHRVA,D55,6760;%%
  
  
  \bibitem{lykken}
  G.~Barenboim, J.~D.~Lykken,
  ``MINOS and CPT-violating neutrinos,''
  Phys.\ Rev.\  {\bf D80}, 113008 (2009).
  [arXiv:0908.2993 [hep-ph]].

  
  
\bibitem{nelson}
  N.~Engelhardt, A.~E.~Nelson, J.~R.~Walsh,
  ``Apparent CPT Violation in Neutrino Oscillation Experiments,''
  Phys.\ Rev.\  {\bf D81}, 113001 (2010).
  [arXiv:1002.4452 [hep-ph]].  


\bibitem{cpt}
  O.~W.~Greenberg,
  ``CPT violation implies violation of Lorentz invariance,''
  Phys.\ Rev.\ Lett.\  {\bf 89}, 231602 (2002).
  [hep-ph/0201258].

\bibitem{gc1}
  N.~Arkani-Hamed, H.~C.~Cheng, M.~A.~Luty and S.~Mukohyama,
  ``Ghost condensation and a consistent infrared modification of gravity,''
  JHEP {\bf 0405}, 074 (2004)
  [arXiv:hep-th/0312099].

\bibitem{gc3}
  T.~Furukawa, S.~Yokoyama, K.~Ichiki, N.~Sugiyama and S.~Mukohyama,
  ``Ghost Dark Matter,''
  JCAP {\bf 1005}, 007 (2010)
  [arXiv:1001.4634 [astro-ph.CO]].

\bibitem{gc2}
  N.~Arkani-Hamed, H.~C.~Cheng, M.~A.~Luty, S.~Mukohyama and T.~Wiseman,
  ``Dynamics of Gravity in a Higgs Phase,''
  JHEP {\bf 0701}, 036 (2007)
  [arXiv:hep-ph/0507120].

\bibitem{hitoshi}
  H.~Murayama and T.~Yanagida,
  ``LSND, SN1987A, and CPT violation,''
  Phys.\ Lett.\  B {\bf 520}, 263 (2001)
  [arXiv:hep-ph/0010178].
  %%CITATION = PHLTA,B520,263;%%


\bibitem{pdg}
K. Nakamura et al. (Particle Data Group), \\
J. Phys. G 37, 075021 (2010)

%\cite{Murayama:2003zw}
\bibitem{Murayama:2003zw}
  H.~Murayama,
  ``CPT tests: Kaon vs neutrinos,''
  Phys.\ Lett.\  B {\bf 597}, 73 (2004)
  [arXiv:hep-ph/0307127].
  %%CITATION = PHLTA,B597,73;%%  
  
\bibitem{nima}
  N.~Arkani-Hamed, S.~Dimopoulos, G.~R.~Dvali {\it et al.},
  ``Neutrino masses from large extra dimensions,''
  Phys.\ Rev.\  {\bf D65}, 024032 (2002).
  [hep-ph/9811448].

\bibitem{neubert}
  Y.~Grossman, M.~Neubert,
  ``Neutrino masses and mixings in nonfactorizable geometry,''
  Phys.\ Lett.\  {\bf B474}, 361-371 (2000).
  [hep-ph/9912408].

\bibitem{park}
  S.~C.~Park, K.~Wang, T.~T.~Yanagida,
  ``Neutrino mass from a hidden world and its phenomenological implications,''
  Phys.\ Lett.\  {\bf B685}, 309-312 (2010).
  [arXiv:0909.2937 [hep-ph]].


%\cite{Randall:1999ee}
\bibitem{rs}
  L.~Randall, R.~Sundrum,
  ``A Large mass hierarchy from a small extra dimension,''
  Phys.\ Rev.\ Lett.\  {\bf 83}, 3370-3373 (1999).
  [hep-ph/9905221].

%\cite{Park:2009cs}
\bibitem{sued}
  S.~C.~Park, J.~Shu,
  ``Split Universal Extra Dimensions and Dark Matter,''
  Phys.\ Rev.\  {\bf D79}, 091702 (2009).
  [arXiv:0901.0720 [hep-ph]].


%\cite{Di Grezia:2005qf}
\bibitem{Di Grezia:2005qf}
  E.~Di Grezia, S.~Esposito and G.~Salesi,
  ``Laboratory bounds on Lorentz symmetry violation in low energy neutrino
  physics,''
  Mod.\ Phys.\ Lett.\  A {\bf 21}, 349 (2006)
  [arXiv:hep-ph/0504245].
  %%CITATION = MPLAE,A21,349;%%

%\cite{ArkaniHamed:2004ar}
\bibitem{ArkaniHamed:2004ar}
  N.~Arkani-Hamed, H.~C.~Cheng, M.~Luty and J.~Thaler,
  ``Universal dynamics of spontaneous Lorentz violation and a new
  spin-dependent inverse-square law force,''
  JHEP {\bf 0507}, 029 (2005)
  [arXiv:hep-ph/0407034].
  %%CITATION = JHEPA,0507,029;%%

%\cite{Grossman:2005ej}
\bibitem{Grossman:2005ej}
  Y.~Grossman, C.~Kilic, J.~Thaler and D.~G.~E.~Walker,
  ``Neutrino constraints on spontaneous Lorentz violation,''
  Phys.\ Rev.\  D {\bf 72}, 125001 (2005)
  [arXiv:hep-ph/0506216].
  %%CITATION = PHRVA,D72,125001;%%
  
  \bibitem{bluebook}
According to the ``Bluebook", Planck's instruments have angular resolution at the level of (5-33) arcmin or $(7.2-47) \times 10^{-4}$ rad. The Bluebook can be downloaded at the official website of Planck project( http://www.rssd.esa.int/index.php?project=planck).
  


\end{thebibliography}
\end{document}